\documentclass[12pt]{JHEP3} 

\usepackage{amsmath,amssymb,axodraw,geometry}
\newcommand{\be}{\begin{equation}}
\newcommand{\ee}{\end{equation}}
\newcommand{\bea}{\begin{eqnarray}}
\newcommand{\eea}{\end{eqnarray}}

\newcommand\fverb{\setbox\pippobox=\hbox\bgroup\verb}
\newcommand\fverbdo{\egroup\medskip\noindent%
                        \fbox{\unhbox\pippobox}\ }
\newcommand\fverbit{\egroup\item[\fbox{\unhbox\pippobox}]}
\newbox\pippobox

\title{Softness of Supersymmetry Breaking 
on the Orbifold $T^2/Z_2$}

\author{ Hyun Min Lee \\
Deutsches Elektronen-Synchrotron DESY, D-22603 Hamburg, Germany \\
E-mail: \email{minlee@mail.desy.de}
}

\preprint{DESY 05-025}      

\abstract{
We consider supersymmetry breaking due to a Scherk-Schwarz twist or localized mass terms
in 6d ${\cal N}=1$ supersymmetric gauge theory compactified
on the orbifold $T^2/Z_2$.
We show that the Scherk-Schwarz breaking in 6d is equivalent to the localized
breaking with mass terms along the lines in extra dimensions.
In the presence of the considered supersymmetry breaking,
we find that there arises a finite one-loop mass correction to a brane scalar
due to the KK modes of bulk gauge fields.
}


\begin{document}

\section{Introduction}


Weak-scale supersymmetry(SUSY) \cite{prnilles} has been a promising candidate for physics beyond the Standard 
Model due to the natural solution to the hierarchy problem and the gauge coupling 
unification and etc.
It is well known that 
supergravity mediation of SUSY breaking at the hidden sector generates 
all required soft SUSY breaking terms of order the weak scale \cite{prnilles}. 
However, it does not explain how soft masses approximately conserve flavor as required
by bounds on flavor-changing neutral currents.

Recently, there has been a lot of attention to models with extra dimensions which give a new
ground for understanding the SUSY breaking in a geometric way. 
Identifying extra dimensions by discrete actions leads to orbifolds \cite{dixon}, 
which lead to chiral fermions and the reduction of higher dimensional supersymmetry. 
Moreover, all or some of SM particles can be regarded to live on the appearing orbifold 
fixed points or branes.


Particularly, one can impose on bulk fields twisted boundary conditions                
in extra dimensions, $\acute{{\it a}}$ {\it la} Scherk-Schwarz(SS) \cite{SS}.       
Then, one can break further the remaining SUSY after orbifolding.
In 5d ${\cal N}=1$ SUSY gauge theory compactified on $S^1/Z_2$, 
it was shown that 
in the presence of the SS breaking of SUSY, there arises a finite one-loop mass correction 
of the zero mode of a bulk scalar or a brane scalar due to the sum of Kaluza-Klein(KK) modes of
bulk fields \cite{one-loop}. It turns out that the SS breaking is equivalent to the case with
a nonzero auxiliary field ($F$ term) of the radion multiplet in the off-shell 
5d supergravity \cite{max,offshellsugra,pomarol}. 
A nonzero twist parameter or $F$ term can be determined dynamically 
after the radion stabilization \cite{chacko,dyndet}.


On the other hand, one can consider the localized breaking of SUSY 
at the orbifold fixed 
points \cite{peskin,nilles2,gaugino,ah,pomarol,weiner,evenmass,nilles,glocal,ckl,nomura,choi-lee}.
For instance, when the remaining SUSY is broken at the hidden brane, only bulk fields
such as gaugino or gravitino get nonzero masses at tree level 
and the broken SUSY is transmitted to the 
visible brane by bulk fields. Then, one can find that the mass spectrum of bulk fields 
and their coupling at the visible brane are equivalent to those in the SS breaking 
without brane mass terms \cite{pomarol,weiner}.
Therefore, there also appears the one-loop finiteness for a scalar mass of the visible 
brane \cite{ah,nomura,choi-lee}, 
which is due to the geometric separation of SUSY breaking from the visible 
brane. If the broken SUSY is mediated dominantly by gaugino, 
the scalar mass becomes flavor-blind which sheds light on the supersymmetric flavor 
problem \cite{gaugino}.


In this paper, we will consider the SUSY breaking in 6d ${\cal N}=1$ 
supersymmetric gauge  
theory compactified on the orbifold $T^2/Z_2$ \cite{lnz}.  
Even if we consider only $U(1)$ gauge group in the bulk, it is straightforward
to extend to the non-abelian gauge group. 
The orbifold fixed points on $T^2/Z_2$ correspond to codimension-two branes. 
First we consider a generalization of the SS breaking of SUSY to the 6d case.
Then, we show that the SS breaking is equivalent to the localized breaking with mass terms 
along the lines rather than points. This localized SUSY breaking
can be realized by positioning the hidden sector at the fixed boundaries
under additional $Z_2$ actions. 
 
For the localized breaking with mass terms at the codimension-two brane, 
however, 
the classical solution of a bulk field 
is singular for an infinitely thin brane \cite{wise}. 
So, one must regularize the zero thickness of brane. 
Then, the regulator dependence in the classical
solution is absorbed into the renormalized brane mass, which has a {\it classical} logarithmic 
RG running \cite{wise}. 
In that sense, the localized breaking at the codimension-two brane is sensitive to
the ultraviolet physics of regularization 
even in a mild way with the log divergence.
Actually, it has been shown \cite{kai} that in the presence of mass terms 
localized at the fixed points, 
the one-loop mass for a brane scalar due to bulk gauge
fields has a log divergence due to the infinitely thin brane.

On the other hand, the localized mass terms at codimension-one branes are 
insensitive to the regularization of the brane thickness, 
as seen from the equivalence to the SS breaking. 
By using the off-shell action for 6d SUSY gauge theory 
with the bulk-brane coupling \cite{lnz}, 
we make a computation of one-loop mass correction to
a brane scalar due to the SS breaking or the localized breaking 
along the distant lines. 
Thus, we find that the resulting one-loop correction is finite.
In the limit of taking one extra dimension without a SS twist to be 
much smaller than the other one with a SS twist, 
we reproduce the 5d result with a SS twist. 
On the other hand, a small extra dimension with 
a nontrivial SS twist is not decoupled but rather its effect is dominant 
in the one-loop mass correction. 

The paper is organized as follows. First we describe the SS twisted boundary conditions on the
bulk gaugino and find the mass spectrum and mode expansion of gaugino.
Next in the localized SUSY breaking with general $Z_2$-even mass terms along the lines, 
we obtain the similar result as in the SS twist. 
Then, we compute the one-loop mass correction 
to a brane scalar due to the KK modes of bulk gauge fields. Finally the conclusion is drawn.

\section{Scherk-Schwarz breaking of SUSY on $T^2/Z_2$}

Let us consider a 6d ${\cal N}=1$ supersymmetric $U(1)$ gauge theory 
compactified on the $T^2/Z_2$ orbifold \footnote{It is straightforward 
to include hypermultiplets coupled to the $U(1)$ \cite{lnz} 
and extend to bulk non-abelian gauge groups. 
In these cases, one needs to remember
that the bulk matter content is severely restricted due to genuine 
6d anomalies \cite{lnz}.}.
Two extra dimensions on a torus are identified as 
$x_5\equiv x_5+2\pi R_5$ and $x_6\equiv x_6+2\pi R_6$ where $R_5$ 
and $R_6$ are radii of extra dimensions. By orbifolding on the torus by $Z_2$, 
we identify $(x_5,x_6)$ with $(-x_5,-x_6)$. 
Then, there appear four orbifold fixed points,
\be
(0,0), \ \ (\pi R_5,0), \ \ (0,\pi R_6), \ \ (\pi R_5,\pi R_6).
\ee 
The fundamental region is the half of the torus.

The kinetic term for the $U(1)$ gaugino \footnote{For notations
and conventions, refer to \cite{lnz}.} is given by
\bea
{\cal L}=i{\bar\Omega}_i\Gamma^M\partial_M\Omega^i.\label{gkin}
\eea
The gaugino $\Omega^i$ is a right-handed simplectic Majorana-Weyl fermion
satisfying the chirality condition
\be
\Gamma^7\Omega^i=\Omega^i.
\ee
On writing the gaugino in a four dimensional Weyl
representation $\Omega^i_R\equiv \lambda^i$, eq.~(\ref{gkin}) becomes 
\bea
{\cal L}=i{\bar\lambda}_i\gamma^M\partial_M\lambda^i. \label{gkina}
\eea

From the symmetry of the action on an orbifold $T^2/Z_2$, let us consider 
the orbifold boundary conditions and the Scherk-Schwarz(SS) twists on $T^2/Z_2$ 
as follows,
\bea
Z_2&:&~~\lambda(x,-x_5,-x_6)=\tau_3(i\gamma^5)\lambda(x,x_5,x_6)
\equiv P\lambda(x,x_5,x_6), \label{orb}\\
T_1&:&~~\lambda(x,x_5+2\pi R_5,x_6)=
U_1\lambda(x,x_5,x_6), \label{tw1}\\
T_2&:&~~\lambda(x,x_5,x_6+2\pi R_6)= 
U_2\lambda(x,x_5,x_6) \label{tw2}
\eea
where $U_i(i=1,2)$ are $2\times 2$ twist matrices corresponding to $SU(2)_R$
rotations.  
The SS twists on the orbifold are subject to the consistency 
conditions $U_iPU_i=P(i=1,2)$ and $U_1U_2=U_2U_1$. 
We note that there is another possible choice of the parity matrix 
$P=\pm {\bf 1}_2(i\gamma^5)$, instead of $P=\tau_3(i\gamma^5)$.
In this case, the consistency conditions lead 
to $U_i=\pm {\bf 1}_2$ or $\pm \tau_3$.
However, in this paper, let us focus on the case with $P=\tau_3(i\gamma^5)$
for which a continuous twist is possible.
 
The first condition $U_iPU_i=P(i=1,2)$ gives rise to
the following form for either $U_1$ or $U_2$:
a continuous twist connected to the identity,
\bea
U_i=e^{-i[2\pi\omega_i(\tau_1\sin\phi_i+\tau_2\cos\phi_i)]}
\eea
with $\omega_i,\phi_i$ being real parameters,
or a discrete twist not connected to the identity,
\bea
U_i=-{\bf 1}_2. 
\eea
By using the residual global invariance, a continuous twist($U_i$ with $i=1$
or $2$) can be always set to the one with $\phi_i=0$.
Therefore, also considering the second condition $U_1U_2=U_2U_1$, 
we find that there are four possible twisted boundary conditions: 
\bea
U_1&=&e^{-2\pi i\omega_5\tau_2}, \ \ U_2=e^{2\pi i\omega_6\tau_2},
\label{twa}\\
U_1&=&-{\bf 1}_2, \ \ U_2=e^{2\pi i\omega_6\tau_2}, \label{twb}\\
U_1&=&e^{-2\pi i\omega_5\tau_2}, \ \ U_2=-{\bf 1}_2, \label{twc}\\
U_1&=&U_2=-{\bf 1}_2 \label{twd}
\eea
where $\omega_5,\omega_6$ are real constant parameters. We note that the 
discrete choice of twist matrices corresponds to using $R$-parity 
of ${\cal N}=1$ 4d supersymmetry as the global symmetry. 

First, for the case with continuous twists in both extra dimensions given by
eq.~(\ref{twa}), 
let us make a redefinition of the gaugino as
\be
\lambda(x,x_5,x_6)=e^{-i(\omega_5x_5/R_5-\omega_6x_6/R_6)\tau_2}
{\tilde\lambda}(x,x_5,x_6).
\ee
Then, regarding $\tilde\lambda$ to be untwisted fields,
we take the redefined gaugino to be a solution to the twisted boundary 
conditions (\ref{tw1}) and (\ref{tw2}) with eq.~(\ref{twa}). 
Moreover, one can show that $\tilde\lambda$ satisfies the same orbifold 
boundary condition as $\lambda$ in eq.~(\ref{orb}). 

Let us write the untwisted fields $\tilde\lambda$ in terms of 4d Majorana
spinors ${\tilde\psi}^i(i=1,2)$ as
\bea
{\tilde\lambda}^1&=&+\frac{1}{2}(1+i\gamma^5){\tilde\psi}^1
+\frac{1}{2}(1-i\gamma^5){\tilde\psi}^2,\\
{\tilde\lambda}^2&=&-\frac{1}{2}(1-i\gamma^5){\tilde\psi}^1
+\frac{1}{2}(1+i\gamma^5){\tilde\psi}^2,
\eea
and similarly for the twisted fields $\lambda$ in terms of 4d Majorana
spinors $\psi^i(i=1,2)$.
We note that the untwisted Majorana spinors are related to the twisted ones by 
\bea
\left(\begin{array}{l} \psi^1 \\ \psi^2 \end{array}\right)
=\left(\begin{array}{ll} \cos\bigg(\frac{\omega_5x_5}{R_5}
+\frac{\omega_6x_6}{R_6}\bigg) & -\sin\bigg(\frac{\omega_5x_5}{R_5}
+\frac{\omega_6x_6}{R_6}\bigg) \\
\sin\bigg(\frac{\omega_5x_5}{R_5}
+\frac{\omega_6x_6}{R_6}\bigg) & \cos\bigg(\frac{\omega_5x_5}{R_5}
+\frac{\omega_6x_6}{R_6}\bigg) \end{array}\right)
\left(\begin{array}{l} {\tilde\psi}^1 \\ {\tilde\psi}^2 \end{array}\right).
\label{majrel}
\eea
Then, from eq.~(\ref{orb}), one can show that 
${\tilde\psi}^i$ satisfy the following $Z_2$ boundary conditions,
\bea
{\tilde\psi}^1(x,-x_5,-x_6)&=&+{\tilde\psi}^1(x,x_5,x_6), \\
{\tilde\psi}^2(x,-x_5,-x_6)&=&-{\tilde\psi}^2(x,x_5,x_6),
\eea
and similarly for $\psi^i$.
With this redefinition of fields, 
let us write the gaugino kinetic term (\ref{gkina}) 
in terms of untwisted fields $\tilde\psi$ as
\bea
{\cal L}&=&i\overline{{\tilde\psi}^1}\gamma^\mu\partial_\mu{\tilde\psi}^1
+i\overline{{\tilde\psi}^2}\gamma^\mu\partial_\mu{\tilde\psi}^2 
-\overline{{\tilde\psi}^1}(\partial_5+\gamma^5\partial_6){\tilde\psi}^2
+\overline{{\tilde\psi}^2}(\partial_5+\gamma^5\partial_6){\tilde\psi}^1 
\nonumber \\
&-&\frac{\omega_5}{R_5}(\overline{{\tilde\psi}^1}{\tilde\psi}^1
+\overline{{\tilde\psi}^2}{\tilde\psi}^2)
+\frac{\omega_6}{R_6}(\overline{{\tilde\psi}^1}\gamma^5{\tilde\psi}^1
+\overline{{\tilde\psi}^2}\gamma^5{\tilde\psi}^2).
\eea
Equivalently, by writing ${\tilde\psi}^i=(\chi^i,{\bar\chi}^i)^T(i=1,2)$ 
with 4d Weyl spinors $\chi^i$, the action becomes
\bea
{\cal L}=&&\sum_{i=1,2}(i\chi^i\sigma^\mu\partial_\mu{\bar\chi}^i
+i{\bar\chi}^i{\bar\sigma}^\mu\partial_\mu\chi^i)
\nonumber \\
&&+[-\chi^1(\partial_5-i\partial_6)\chi^2
+\chi^2(\partial_5-i\partial_6)\chi^1+c.c.]
+{\cal L}_m \label{fnl}
\eea
where ${\cal L}_m$ corresponds to the bulk mass terms given by
\bea
{\cal L}_m=-\bigg[\bigg(\frac{\omega_5}{R_5}+i\frac{\omega_6}{R_6}\bigg)
(\chi^1\chi^1+\chi^2\chi^2)+c.c.\bigg].\label{twistedmass}
\eea
Therefore, we find that 
the SS twist on the torus induces nonzero $Z_2$-even mass terms  
in the basis of the untwisted gaugino that we have introduced 
for redefining the gaugino. 
We note that the 
$Z_2$-even and odd untwisted fields take equal bulk masses as in the 5d case.  

From the action (\ref{fnl}), we can derive 
the equations of motion for gaugino as follows, 
\bea
i\sigma^\mu\partial_\mu{\bar\chi}^2+(\partial_5-i\partial_6)\chi^1
-\bigg(\frac{\omega_5}{R_5}+i\frac{\omega_6}{R_6}\bigg)\chi^2&=&0, \\
i{\bar\sigma}^\mu\partial_\mu\chi^1-(\partial_5+i\partial_6){\bar\chi}^2
-\bigg(\frac{\omega_5}{R_5}-i\frac{\omega_6}{R_6}\bigg){\bar\chi}^1&=&0.
\eea
Therefore, solving the above equations, we find the solution for the untwisted 
gaugino as 
\be
\left(\begin{array}{l}\chi^1 \\ \chi^2\end{array}\right)(x,x_5,x_6)
=\frac{1}{2\pi\sqrt{R_5R_6}}
\sum_{n_5,n_6\in {\bf Z}} 
\left(\begin{array}{l} \cos\bigg(\frac{n_5}{R_5}x_5-\frac{n_6}{R_6}x_6\bigg) 
\\ \sin\bigg(\frac{n_5}{R_5}x_5-\frac{n_6}{R_6}x_6\bigg) \end{array}\right)
\eta^{(n_5,n_6)}(x)
\ee
where $n_5,n_6$ are integer, 
$i\sigma^\mu\partial_\mu{\bar\eta}^{(n_5,n_6)}(x)=M_{n_5,n_6}\eta^{(n_5,n_6)}(x)$ 
and the mass spectrum is given by
\be
M_{n_5,n_6}=\frac{n_5+\omega_5}{R_5}+i\bigg(\frac{n_6+\omega_6}{R_6}\bigg).
\label{ssmass}
\ee
Consequently, due to the relation (\ref{majrel}), 
the solution for the twisted gaugino $\psi^i=(\zeta^i,{\bar\zeta}^i)^T(i=1,2)$ 
becomes
\be
\left(\begin{array}{l}\zeta^1 \\ \zeta^2\end{array}\right)(x,x_5,x_6)
=\frac{1}{2\pi\sqrt{R_5R_6}}
\sum_{n_5,n_6\in {\bf Z}}
\left(\begin{array}{l} \cos\bigg(\frac{n_5+\omega_5}{R_5}x_5-\frac{n_6+\omega_6}{R_6}x_6\bigg)
\\ \sin\bigg(\frac{n_5+\omega_5}{R_5}x_5-\frac{n_6+\omega_6}{R_6}x_6\bigg) 
\end{array}\right)
\eta^{(n_5,n_6)}(x).\label{twistedg}
\ee

Similarly, for the case with a continuous twist in one direction
and a discrete twist in the other direction given by eq.~(\ref{twb}),
we can make a redefinition of the gaugino 
with ${\tilde\lambda}$ as
\be
\lambda(x,x_5,x_6)=e^{i(\omega_6x_6/R_6)\tau_2}{\tilde\lambda}(x,x_5,x_6).
\ee  
Then, ${\tilde\lambda}$ satisfies the following orbifold 
and twisted boundary conditions:
\bea
Z_2&:&~~{\tilde\lambda}(x,-x_5,-x_6)=\tau_3(i\gamma^5){\tilde\lambda}(x,x_5,x_6)
\label{orba}\\
T_1&:&~~{\tilde\lambda}(x,x_5+2\pi R_5,x_6)=- {\tilde\lambda}(x,x_5,x_6), \label{tw1a}\\
T_2&:&~~{\tilde\lambda}(x,x_5,x_6+2\pi R_6)= {\tilde\lambda}(x,x_5,x_6). 
\label{tw2a}
\eea
Consequently, plugging the redefined gaugino into the action, 
deriving the equation for ${\tilde\lambda}$ 
and imposing the above boundary conditions to ${\tilde\lambda}$, 
we find the corresponding solution for ${\tilde\lambda}$  
in 4d Weyl representation as
\be
\left(\begin{array}{l}\chi^1 \\ \chi^2\end{array}\right)(x,x_5,x_6)
=\frac{1}{2\pi\sqrt{R_5R_6}}
\sum_{n_5,n_6\in {\bf Z}}
\left(\begin{array}{l} \cos\bigg(\frac{n_5+\frac{1}{2}}{R_5}x_5-\frac{n_6}{R_6}x_6\bigg)
\\ \sin\bigg(\frac{n_5+\frac{1}{2}}{R_5}x_5-\frac{n_6}{R_6}x_6\bigg) 
\end{array}\right)
\eta^{(n_5,n_6)}(x)\label{untwisted}
\ee
where $n_5,n_6$ are integer,
$i\sigma^\mu\partial_\mu{\bar\eta}^{(n_5,n_6)}(x)=M_{n_5,n_6}\eta^{(n_5,n_6)}(x)$ 
and the mass spectrum is given by
\be
M_{n_5,n_6}=\frac{n_5+\frac{1}{2}}{R_5}+i\bigg(\frac{n_6+\omega_6}{R_6}\bigg).
\label{ssmassa}
\ee
Therefore, the solution for the twisted gaugino $\lambda$  
is given by eq.~(\ref{twistedg}) with $\omega_5=\frac{1}{2}$.
Also for the case with twist matrices (\ref{twc}),
we only have to interchange $(n_5,R_5)\leftrightarrow (n_6,R_6)$
with $\omega_6\rightarrow\omega_5$ in eq.~(\ref{untwisted}), and then obtain
the solution for the twisted gaugino $\lambda$ given by eq.~(\ref{twistedg}) 
with $\omega_6=\frac{1}{2}$.

Lastly, for the case with discrete twists in both extra dimensions, 
the solution for the twisted gaugino $\lambda$ 
is given by eq.~(\ref{twistedg}) with $\omega_5=\omega_6=\frac{1}{2}$.

\section{SUSY breaking due to localized gaugino masses}

In this section, instead of the SS boundary twists of gaugino, 
let us consider a local breaking of supersymmetry which
is parametrized by gaugino mass terms, and show the equivalence between the SS
breaking and the localized breaking. 

Let us take the most general $Z_2$-even mass 
terms \footnote{If one introduces
gaugino mass terms proportional to $\delta(x_5)$ and $\delta(x_6)$, 
there appears a non-supersymmetric gauge coupling at the origin 
due to the suppression of gaugino wave function \cite{choi-lee}. 
Since we assume the visible sector fields to be localized at the origin, 
let us consider the gaugino mass terms only at distant lines.} for gaugino,
which are localized along the two lines intersecting at a fixed point 
$(\pi R_5,\pi R_6)$ on the orbifold,  
\bea
{\cal L}_m&=&-[2m(\chi^1\chi^1+\rho\chi^2\chi^2)+c.c]\delta(x_5-\pi R_5) 
\nonumber \\
&&-[2im'(\chi^1\chi^1+\rho'\chi^2\chi^2)+c.c.]\delta(x_6-\pi R_6)
\eea
where $(m,\rho)$ and $(m',\rho')$ are gaugino mass parameters and 
they are assumed to be real. 
Then, the lines with localized mass terms should be 
regarded as the fixed boundaries under two additional independent $Z_2$ 
actions \cite{kkl}:
$Z'_2$: $(x_5,x_6)\rightarrow (-x_5,x_6)$ and $Z^{\prime\prime}_2$: 
$(x_5,x_6)\rightarrow (x_5,-x_6)$. In this case, it is conceivable that 
the localized mass terms 
are due to the SUSY breaking in the hidden sector located on the lines,
rather than points. 

In this case, the gaugino equations of motion are
\bea
i\sigma^\mu\partial_\mu{\bar\chi}^2+(\partial_5-i\partial_6)\chi^1
-2(m\rho\delta(x_5-\pi R_5)+im'\rho'\delta(x_6-\pi R_6))\chi^2&=&0, \label{geq1}\\
i{\bar\sigma}^\mu\partial_\mu\chi^1-(\partial_5+i\partial_6){\bar\chi}^2
-2(m\delta(x_5-\pi R_5)-im'\delta(x_6-\pi R_6)){\bar \chi}^1&=&0.\label{geq2}
\eea

Now let us take the solution of gaugino to the above equations as
\be
\left(\begin{array}{l}\chi^1 \\ \chi^2\end{array}\right)(x,x_5,x_6)
= \sum_{M}N_M \left(\begin{array}{l} u^1(x_5,x_6) 
\\ u^2(x_5,x_6) \end{array}\right)
\eta_M(x)
\ee
where $N_M$ is the normalization constant and $i\sigma^\mu\partial_\mu{\bar\eta}_M(x)=M\eta_M(x)$. Then, the gaugino equations are
\bea
M{\bar u}^2+(\partial_5-i\partial_6)u^1-2(m\rho\delta(x_5-\pi R_5)
+im'\rho'\delta(x_6-\pi R_6))u^2&=&0,\label{geq1a}\\
{\bar M}u^1-(\partial_5+i\partial_6){\bar u}^2-2(m\delta(x_5-\pi R_5)
-im'\delta(x_6-\pi R_6)){\bar u}^1&=&0.\label{geq2a}
\eea
Let us take $u^1,u^2$ to be real functions. 
Then, taking $M=M_5+iM_6$ with real $M_5$ and $M_6$ and 
using eqs.~(\ref{geq1a}) and (\ref{geq2a}), 
we obtain the equation for $t\equiv u^2/u^1$ as
\bea
\partial_5 t&=&M_5(1+t^2)-2m(1+\rho t^2)\delta(x_5-\pi R_5), \\
\partial_6 t&=&-M_6(1+t^2)+2m'(1+\rho' t^2)\delta(x_6-\pi R_6).
\eea
It is convenient to consider the $Z_2$-odd solution of $t$ separately 
around different fixed points and match them in the overlap 
regions \cite{choi-lee}.
That is, let us consider the solution of $t$ which satisfies the equations of
motion inside a torus centered at each fixed point.
Thus, we find the solution for $t$: 
\begin{itemize}
\item $-\pi R_5<x_5<\pi R_5$ and $-\pi R_6<x_6<\pi R_6$,
\be
t=\tan(M_5x_5-M_6x_6).
\ee 

\item $0<x_5<2\pi R_5$ and $-\pi R_6<x_6<\pi R_6$,
\be
t=\tan[M_5(x_5-\pi R_5)-M_6x_6-{\rm arctan}\alpha(\rho,m\epsilon(x_5-\pi R_5))]
\ee
where $\epsilon(x_5-\pi R_5)$ is a step function with $2\pi R_5$ periodicity 
given by
\be
\epsilon(x_5)=\left\{\begin{array}{l} +1, \ \ 0<x_5<\pi R_5, \\
0, \ \ x_5=0, \\ -1, \ \ -\pi R_5<x_5<0,\end{array}\right.
\ee
and
\be
\alpha(\rho,m\epsilon(x_5-\pi R_5))\equiv
\frac{1}{\sqrt{\rho}}\tan(\sqrt{\rho}m\epsilon(x_5-\pi R_5)).
\ee
\item $-\pi R_5<x_5<\pi R_5$ and $0<x_6<2\pi R_6$,
\be
t=\tan[M_5x_5-M_6(x_6-\pi R_6)
+{\rm arctan}\alpha(\rho',m'{\tilde\epsilon}(x_6-\pi R_6))],
\ee
where ${\tilde\epsilon}(x_6-\pi R_6)$ is a step function 
with $2\pi R_6$ periodicity.
\item $0<x_5<2\pi R_5$ and $0<x_6<2\pi R_6$,
\bea
t&=&\tan[M_5(x_5-\pi R_5)-M_6(x_6-\pi R_6)
-{\rm arctan}\alpha(\rho,m\epsilon(x_5-\pi R_5)) \nonumber \\
&+&{\rm arctan}\alpha(\rho',m'{\tilde \epsilon}(x_6-\pi R_6))].
\eea

\end{itemize}
Identify the first two solutions in the overlap region 
of $0<x_5<\pi R_5$ and $-\pi R_6<x_6<\pi R_6$, 
we find 
\be
M_5=\frac{1}{R_5}\bigg(n_5+\frac{1}{\pi}{\rm arctan}\alpha(\rho,m)\bigg),
\ \ \  n_5={\rm integer}. 
\ee
Likewise, identifying the first and third solutions in the overlap region
of $-\pi R_5<x_5<\pi R_5$ and $0<x_6<\pi R_6$, we also find
\be
M_6=\frac{1}{R_6}\bigg(n_6+\frac{1}{\pi}{\rm arctan}\alpha(\rho',m')
\bigg), \ \ \ n_6={\rm integer}. 
\ee 
Then, comparing the other solutions in the overlap regions 
does not lead to a new condition. Therefore, the mass spectrum is equivalent to
the one with SS breaking when $\omega_5$ and $\omega_6$ in eq.~(\ref{ssmass})
are identified with ${\rm arctan}\alpha(\rho,m)/\pi$ 
and ${\rm arctan}\alpha(\rho',m')/\pi$, respectively.

Moreover, the solutions of $u^1$ and $u^2$ are also given in the separate 
regions:
\begin{itemize}
\item $-\pi R_5<x_5<\pi R_5$ and $-\pi R_6<x_6<\pi R_6$,
\bea
u^1&=&\cos(M_5x_5-M_6x_6), \label{sol11}\\
u^2&=&\sin(M_5x_5-M_6x_6). 
\eea

\item $0<x_5<2\pi R_5$ and $-\pi R_6<x_6<\pi R_6$,
\bea
u^1&=&(-1)^{n_5}A(\rho,m\epsilon(x_5-\pi R_5))\times \nonumber \\
&\times&\cos[M_5(x_5-\pi R_5)
-M_6x_6-{\rm arctan}\alpha(\rho,m\epsilon(x_5-\pi R_5))], \\
u^2&=&(-1)^{n_5}A(\rho,m\epsilon(x_5-\pi R_5))\times \nonumber \\
&\times&\sin[M_5(x_5-\pi R_5)
-M_6x_6-{\rm arctan}\alpha(\rho,m\epsilon(x_5-\pi R_5))]
\eea
where
\be
A(\rho,m\epsilon(x_5-\pi R_5))\equiv 
\bigg(\frac{1+\alpha^2(\rho,m\epsilon(x_5-\pi R_5))}
{1+\rho\alpha^2(\rho,m\epsilon(x_5-\pi R_5))}\bigg)^{1/2}.
\ee
\item $-\pi R_5<x_5<\pi R_5$ and $0<x_6<2\pi R_6$,
\bea
u^1&=&(-1)^{n_6}A(\rho',m'{\tilde\epsilon}(x_6-\pi R_6))\times \nonumber \\
&\times&\cos[M_5x_5-M_6(x_6-\pi R_6)
+{\rm arctan}\alpha(\rho',m'{\tilde\epsilon}(x_6-\pi R_6))],\\
u^2&=&(-1)^{n_6}A(\rho',m'{\tilde\epsilon}(x_6-\pi R_6))\times \nonumber \\
&\times&\sin[M_5x_5-M_6(x_6-\pi R_6)
+{\rm arctan}\alpha(\rho',m'{\tilde\epsilon}(x_6-\pi R_6))].
\eea

\item $0<x_5<2\pi R_5$ and $0<x_6<2\pi R_6$,
\bea
u^1&=&(-1)^{n_5+n_6}
A(\rho,m\epsilon(x_5-\pi R_5))A(\rho',m'{\tilde\epsilon}(x_6-\pi R_6))
\times \nonumber \\
&\times&\cos[M_5(x_5-\pi R_5)-M_6(x_6-\pi R_6)\nonumber \\
&-&{\rm arctan}\alpha(\rho,m\epsilon(x_5-\pi R_5))
+{\rm arctan}\alpha(\rho',m'{\tilde\epsilon}(x_6-\pi R_6))],\\
u^2&=&(-1)^{n_5+n_6}
A(\rho,m\epsilon(x_5-\pi R_5))A(\rho',m'{\tilde\epsilon}(x_6-\pi R_6))
\times \nonumber \\
&\times&\sin[M_5(x_5-\pi R_5)-M_6(x_6-\pi R_6)\nonumber \\
&-&{\rm arctan}\alpha(\rho,m\epsilon(x_5-\pi R_5))
+{\rm arctan}\alpha(\rho',m'{\tilde\epsilon}(x_6-\pi R_6))].
\eea

\end{itemize}

In order to make a normalization of KK modes, let us insert the solutions
in the action and integrate it over extra dimensions. Then, we obtain the
normalization constant in the separate regions:
\begin{itemize}
\item $-\pi R_5<x_5<\pi R_5$ and $-\pi R_6<x_6<\pi R_6$,
\be
N_M=\bigg(\int_{-\pi R_5}^{\pi R_5}dx_5\int_{-\pi R_6}^{\pi R_6}dx_6 
[(u^1)^2+(u^2)^2]\bigg)^{-1/2}=\frac{1}{2\pi\sqrt{R_5R_6}}.\label{norm1}
\ee

\item $0<x_5<2\pi R_5$ and $-\pi R_6<x_6<\pi R_6$,
\be
N_M=\bigg(\int_0^{2\pi R_5}dx_5\int_{-\pi R_6}^{\pi R_6}dx_6
[(u^1)^2+(u^2)^2]\bigg)^{-1/2}=\frac{1}{2\pi\sqrt{R_5R_6}}A^{-1}(\rho,m).
\ee

\item $-\pi R_5<x_5<\pi R_5$ and $0<x_6<2\pi R_6$,
\be
N_M=\bigg(\int_{-\pi R_5}^{\pi R_5}dx_5\int_0^{2\pi R_6}dx_6
[(u^1)^2+(u^2)^2]\bigg)^{-1/2}=\frac{1}{2\pi\sqrt{R_5R_6}}A^{-1}(\rho',m').
\ee

\item $0<x_5<2\pi R_5$ and $0<x_6<2\pi R_6$,
\bea
N_M&=&\bigg(\int_0^{2\pi R_5}dx_5\int_0^{2\pi R_6}dx_6
[(u^1)^2+(u^2)^2]\bigg)^{-1/2}\nonumber \\
&=&\frac{1}{2\pi\sqrt{R_5R_6}}
A^{-1}(\rho,m)A^{-1}(\rho',m').
\eea

\end{itemize}

\section{One-loop mass correction to a brane scalar}

In the general case with nonzero gaugino masses, let us put a chiral multiplet 
at the $(0,0)$ fixed point. Then, the scalar partner of the chiral multiplet 
does not feel the supersymmetry breaking directly but there exists a loop 
contribution to its mass due to the distant supersymmetry breaking.
Only $Z_2$-even gaugino couples to the brane scalar.
From the solution (\ref{sol11}) with normalization (\ref{norm1}) in the region
$-\pi R_5<0<\pi R_5$ and $-\pi R_6<x_6<\pi R_6$,
we find that all KK modes of $Z_2$-even gaugino have the same brane 
coupling as the one of bulk gauge boson,
\be
g_4=\frac{g_6}{2\pi\sqrt{R_5R_6} }
\ee
where $g_6$ is the six-dimensional gauge coupling. One has the KK mass 
spectrums for gauge bosons and gaugino running in loops, respectively,
\bea
M^2_{(0)n_5,n_6}&=&\bigg(\frac{n_5}{R_5}\bigg)^2
+\bigg(\frac{n_6}{R_6}\bigg)^2, \\
M^2_{n_5,n_6}&=&\bigg(\frac{n_5+\omega_5}{R_5}\bigg)^2
+\bigg(\frac{n_6+\omega_6}{R_6}\bigg)^2.
\eea

On the other hand,
from the even-mode $\zeta^1$ from eq.~(\ref{twistedg})
at the $(0,0)$ fixed point
and the mass spectrum in eq.~(\ref{ssmass}),  
one can find that the SS breaking leads to the same brane coupling  
and mass spectrum of gaugino as in the localized breaking of supersymmetry.
So, the brane scalar fields do not feel the difference between the SS twist 
and the localized gaugino masses along the distant lines.  

Now let us consider the KK mode contribution to the one-loop mass correction
for a brane scalar $\phi$ with charge $Q$ under the $U(1)$. For this, we note that the coupling
of the bulk auxiliary field to the brane scalar is given by the following action \cite{lnz},
\bea
\int d^6 x\bigg[\frac{1}{2}(D^3)^2+\delta(x^5)\delta(x^6)g_6Q\phi^\dagger 
(-D^3+F_{56})\phi\bigg]
\eea 
where $D^3$ is the third component of auxiliary field in the bulk vector 
multiplet
and $F_{56}$ is the extra component of field strength.
After eliminating the auxiliary field by its equation of motion, we find the resulting 
coupling as
\bea
\int d^4 x\bigg[-g_6Q\phi^\dagger F_{56}(x,x_5=0,x_6=0)\phi
-\frac{1}{2}g^2_6Q^2(\phi^\dagger\phi)^2 \delta(0)\delta(0)\bigg]
\eea
with
\bea
\delta(0)\delta(0)&=&\frac{1}{4\pi^2R_5R_6}\sum_{n_5,n_6\in {\bf Z}}1 \nonumber \\
&=&\frac{1}{4\pi^2R_5R_6}\sum_{n_5,n_6\in {\bf Z}}
\frac{p^2-M^2_{(0)n_5,n_6}}{p^2-M^2_{(0)n_5,n_6}}.
\eea 
Therefore, considering the similar Feynman diagrams as in 5d \cite{peskin,ckl},  
in the dimensional regularization with $d=4-\epsilon$, 
bosonic and fermionic loop contributions to the scalar self energy 
are, at nonzero external momentum $q^2$, respectively, 
\bea
-im^2_B(q^2)=4g^2_4 Q^2\mu^{4-d}\sum_{n_5,n_6\in {\bf Z}}\int 
\frac{d^d p}{(2\pi)^d}\frac{p(q+p)}{(p^2-M^2_{(0)n_5,n_6})(q+p)^2} 
\eea
and
\bea
-im^2_F(q^2)=-4g^2_4 Q^2\mu^{4-d}\sum_{n_5,n_6\in {\bf Z}}\int
\frac{d^d p}{(2\pi)^d}\frac{p(q+p)}{(p^2-M^2_{n_5,n_6})(q+p)^2} 
\eea
By using the Schwinger representation
\bea
\frac{1}{A^n}=\frac{1}{\Gamma(n)}\int^\infty_0 dt\, t^{n-1} e^{-At},
\eea
and performing the momentum integrations via the identities
\bea
\int^\infty_0 dy \,y^{2n+d-1}e^{-y^2t}=\frac{\Gamma(d/2+n)}{2t^{d/2+n}},
\eea  
we find the one-loop corrections as
\bea
m^2_B(q^2)=\frac{g^2_4Q^2(\mu\pi R_5)^\epsilon}{4\pi^3R^2_5}\int^1_0 dx
\bigg[(2-\frac{\epsilon}{2}){\cal J}_2[0,0,c]
+\pi x(1-x)q^2 R^2_5{\cal J}_1[0,0,c]\bigg]
\eea
and
\bea
m^2_F(q^2)&=&-\frac{g^2_4Q^2(\mu \pi R_5)^\epsilon}{4\pi^3R^2_5}\int^1_0 dx
\bigg[(2-\frac{\epsilon}{2}){\cal J}_2[\omega_5,\omega_6,c] \nonumber \\ 
&&+\pi x(1-x)q^2 R^2_5{\cal J}_1[\omega_5,\omega_6,c]\bigg]
\eea
with
\bea
&&{\cal J}_j[\omega_5,\omega_6,c]\equiv \sum_{n_5,n_6\in {\bf Z}}
\int^\infty_0 \frac{dt}{t^{j-\epsilon/2}}
e^{-\pi t[c+a_5(n_5+\omega_5)^2+a_6(n_6+\omega_6)^2]}, \ \ j=1,2; \ \
a_{5,6},c>0; \nonumber \\
&&a_5\equiv x, \ \ a_6\equiv x\bigg(\frac{R_5}{R_6}\bigg)^2, \ \
c\equiv -x(1-x)q^2R^2_5.
\eea

For small positive\footnote{$c$ is positive after a Wick rotation
$q^2=-q^2_E$.} $c$, 
we obtain the following approximate formulas \cite{ghilencea}
for ${\cal J}_j[\omega_5,\omega_6,c]$,
\bea
{\cal J}_1[\omega_5,\omega_6,c\ll 1]&\simeq& \frac{\pi c}{\sqrt{a_5a_6}}
\bigg[\frac{-2}{\epsilon}\bigg]
-\ln\bigg|\frac{\vartheta_1(\omega_6-iu\omega_5|iu)}{(\omega_6-iu\omega_5)\eta(iu)} 
e^{-\pi u\omega^2_5}\bigg|^2 \nonumber \\
&&-\ln[(c+a_5\omega^2_5+a_6\omega^2_6)/a_6], \ \ \ \ 
u\equiv \sqrt{\frac{a_5}{a_6}},\\ 
{\cal J}_2[\omega_5,\omega_6,c\ll 1]&\simeq& -\frac{\pi^2c^2}{2\sqrt{a_5a_6}}
\bigg[\frac{-2}{\epsilon}\bigg]
+\frac{\pi^2a_5}{3}\bigg(\frac{a_5}{a_6}\bigg)^{1/2}
\bigg[\frac{1}{15}-2\Delta^2_{\omega_5}(1-\Delta_{\omega_5})^2\bigg] 
\nonumber \\ 
&&+\bigg[\sqrt{a_5a_6}\sum_{n\in {\bf Z}}|n+\omega_5|{\rm Li}_2(e^{-2\pi iz})
+c.c.\bigg] \nonumber \\
&&+\bigg[\frac{a_6}{2\pi}\sum_{n\in {\bf Z}}{\rm Li}_3(e^{-2\pi iz})+c.c.\bigg]
\eea
where $\Delta_{\omega_5}\equiv\omega_5-[\omega_5]$ 
with $0\leq \Delta_{\omega_5}<1$ and $[\omega_5]\in {\bf Z}$, 
and $z\equiv \omega_6-i\sqrt{\frac{a_5}{a_6}}|n+\omega_5|$.
Here, $\vartheta_1$ is the Jacobi theta function and $\eta$ is the Dedekind
eta function. And ${\rm Li}_2,{\rm Li}_3$ are the polylogarithm functions as
\be
{\rm Li}_n(x)=\sum_{k=1}^\infty\frac{x^k}{k^n}, \ \ n=2,3.
\ee

Therefore, the resulting one-loop correction for the brane scalar 
is given by
\bea
m^2_\phi(q^2)&=&m^2_B(q^2)+m^2_F(q^2) \nonumber \\
&=&\frac{g^2_4Q^2}{2\pi^3R^2_5}\int^1_0 dx\bigg[{\cal J}_2[0,0,c]
-{\cal J}_2[\omega_5,\omega_6,c]\bigg] \nonumber \\
&&+\frac{g^2_4Q^2}{4\pi^2 R^2_5}(q^2R^2_5)\int^1_0 dx\, x(1-x)
\bigg[{\cal J}_1[0,0,c]-{\cal J}_1[\omega_5,\omega_6,c]\bigg].
\eea
Consequently, we observe that both divergences of ${\cal J}_1$ and ${\cal J}_2$
are cancelled and there appear only finite corrections. 
Thus, we can take $c=0$
safely at the zero external momentum without involving the UV and IR mixing 
found in \cite{ghilencea}. So, the mass correction with $q^2=0$ is given by
\bea
m^2_\phi(0)&=&\frac{g^2_4Q^2}{4\pi^3R^2_5}\bigg[\frac{2\pi^2}{3}r
\Delta^2_{\omega_5}(1-\Delta_{\omega_5})^2 
+\frac{1}{r}(I_1(0,0)-I_1(\omega_5,\omega_6)+c.c.) \nonumber
\\
&&+\frac{1}{2\pi r^2}(I_2(0,0)-I_2(\omega_5,\omega_6)
+c.c.)\bigg], \ \ \ r\equiv \frac{R_6}{R_5} \label{1loopmass}
\eea
with 
\bea
I_1(\omega_5,\omega_6)\equiv
\sum_{n\in {\bf Z}}|n+\omega_5|{\rm Li}_2(e^{-2\pi |n+\omega_5|r
-2\pi i\omega_6}), 
\eea
and 
\bea
I_2(\omega_5,\omega_6)\equiv
\sum_{n\in {\bf Z}}{\rm Li}_3(e^{-2\pi |n+\omega_5|r-2\pi i\omega_6}).
\eea

In order to see the mass correction explicitly, let us simplify the sums as follows, 
\bea
I_1(\omega_5,\omega_6)
&=&\frac{1}{2}\sum_{k=1}^{\infty}
\frac{e^{-2\pi i k\Delta_{\omega_6}}}{k^2{\rm sinh}^2(\pi k r)}
\bigg[\Delta_{\omega_5}{\rm cosh}(2\pi k(1-\Delta_{\omega_5})r)
\nonumber \\
&&+(1-\Delta_{\omega_5}){\rm cosh}(2\pi k \Delta_{\omega_5}r)\bigg],
\eea
and
\bea
I_2(\omega_5,\omega_6)
=\sum_{k=1}^{\infty}\frac{e^{-2\pi i k\Delta_{\omega_6}}}{k^3}
\frac{{\rm cosh}(\pi k(1-2\Delta_{\omega_5})r)}{{\rm sinh}(\pi k r)}.
\eea 
Here $\Delta_{\omega_6}\equiv\omega_6-[\omega_6]$ 
with $0\leq \Delta_{\omega_6}<1$ and $[\omega_6]\in {\bf Z}$.
Therefore, inserting the above expressions into eq.~(\ref{1loopmass}),
we find that the resulting mass correction is finite as 
\bea
m^2_\phi(0)&=&\frac{g^2_4Q^2}{4\pi^3R^2_5}\bigg[\frac{2\pi^2}{3}
r\Delta^2_{\omega_5}(1-\Delta_{\omega_5})^2 \nonumber \\
&&+\frac{1}{r}\sum_{k=1}^{\infty}
\frac{1}{k^2{\rm sinh}^2(\pi k r)}\bigg(1-\cos(2\pi k\Delta_{\omega_6})
\{\Delta_{\omega_5}{\rm cosh}(2\pi k(1-\Delta_{\omega_5})r) \nonumber \\
&&+(1-\Delta_{\omega_5}){\rm cosh}(2\pi k \Delta_{\omega_5}r)\}\bigg)
\nonumber \\
&&+\frac{1}{\pi r^2}
\sum_{k=1}^{\infty}
\frac{1}{k^3{\rm tanh}(\pi k r)}\bigg(1-\cos(2\pi k\Delta_{\omega_6})
\frac{{\rm cosh}(\pi k(1-2\Delta_{\omega_5})r)}{{\rm cosh}(\pi kr)}
\bigg)\bigg].\label{massfinal}
\eea 

First let us consider the case with $\Delta_{\omega_5}=0$. 
Then, eq.~(\ref{massfinal}) becomes
\bea
m^2_\phi(0)&=&\frac{g^2_4Q^2}{4\pi^3R^2_5}\bigg[\frac{1}{r}\sum_{k=1}^{\infty}
\frac{1}{k^2{\rm sinh}^2(\pi k r)}(1-\cos(2\pi k\Delta_{\omega_6})) \nonumber \\
&&+\frac{1}{\pi r^2}
\sum_{k=1}^{\infty}
\frac{1}{k^3{\rm tanh}(\pi k r)}(1-\cos(2\pi k\Delta_{\omega_6}))\bigg].
\eea
In this case, let us take the limit of $\pi r\gg 1$, i.e. 
one extra dimension with radius $R_5$ to be much smaller than the other. 
Thus, the resulting mass correction reproduces exactly the 5d case 
with a SS twist \cite{choi-lee},
\bea
m^2_\phi(0)\simeq \frac{g^2_4Q^2}{4\pi^4R^2_6}\sum_{k=1}^{\infty}
\frac{1}{k^3}(1-\cos(2\pi k\Delta_{\omega_6})).
\eea

On the other hand, when one takes $\Delta_{\omega_5}=\frac{1}{2}$, 
which is the case with a discrete twist in the fifth direction,
eq.~(\ref{massfinal}) becomes
\bea
m^2_\phi(0)&=&\frac{g^2_4Q^2}{4\pi^3R^2_5}\bigg[\frac{\pi^2}{24}r+\frac{1}{r}\sum_{k=1}^{\infty}
\frac{1}{k^2{\rm sinh}^2(\pi k r)} \nonumber \\
&&+\frac{1}{\pi r^2}\sum_{k=1}^{\infty}\frac{1}{k^3{\rm tanh}(\pi k r)}
\bigg(1-\cos(2\pi k\Delta_{\omega_6})\frac{\pi k r}{{\rm sinh}(\pi k r)}\bigg)\bigg].
\eea
Again in the limit of $\pi r\gg 1$, the resulting mass correction is
\bea
m^2_\phi(0)\simeq \frac{g^2_4Q^2}{96\pi R^2_5}\bigg(\frac{R_6}{R_5}\bigg)
\bigg[1+{\cal O}\bigg(\frac{R^2_5}{R^2_6}\bigg)\bigg].
\eea
Therefore, in this case, one extra dimension with small radius $R_5$ is not decoupled, 
but rather the effect due to the nontrivial SS twist in that direction is a dominant
contribution to the mass correction.
For other nonzero values of $\Delta_{\omega_5}$, such a non-decoupling of small extra dimension 
remains true because the first term in eq.~(\ref{massfinal}) is dominant for $\pi r\gg 1$.

\section{Conclusion}
We considered supersymmetry breaking on the orbifold $T^2/Z_2$ via the SS 
twisted boundary conditions or the localized mass terms.
It turns out that the SS breaking is equivalent to the localized breaking
at the lines which should be regarded to be fixed boundaries 
under additional $Z_2$ actions. 
In this case, we have shown that in the presence of the SS twist 
or localized mass terms for the bulk gauge sector, 
there arises a finite one-loop mass correction to the visible brane scalar. 
In particular, for the case with one extra 
dimension much smaller than the other, 
we observe that the effect from the small
extra dimension to the one-loop mass correction is not decoupled 
due to a nontrivial SS twist in that direction. 

In order to know whether the contribution due to the bulk gaugino 
dominates over
other contributions such as anomaly mediation \cite{chacko}, 
one needs to determine the SS twist parameter dynamically.
At the level of 4d effective supergravity, 
one could think of the SS breaking to be equivalent to a nonzero $F$ term 
of the corresponding radion multiplet for two extra dimensions 
as in 5d case \cite{chacko},
and introduce a radius stabilization mechanism to determine the $F$ term 
dynamically. Moreover, in order to estimate supergravity loop corrections 
as in 5d case \cite{sugra}, it seems indispensible to understand the 6d 
off-shell supergravity, which is not available yet. Let us leave these
issues in a future publication.

\section*{Acknowledgements}
The author would like to thank W. Buchm$\ddot{{\rm u}}$ller, A. Falkowski 
and D. Ghilencea for useful comments and discussion.

\end{document}